\documentclass[aps, prd, floats, floatfix, twocolumn, nofootinbib, superscriptaddress]{revtex4-1}
\usepackage[pdftex]{graphicx}
\usepackage{color}
\usepackage{latexsym}
\usepackage[colorlinks]{hyperref}
\usepackage{amsthm,amsmath,amssymb}

\newcommand{\be}{\begin{eqnarray}}
\newcommand{\ee}{\end{eqnarray}}
\newcommand{\rar}{\rightarrow}

\begin{document}

\title{Testing Black Holes with Interstellar Missions: I. Orbiting Probes}

\author{Leda~Gao}
\affiliation{Center for Astronomy and Astrophysics, Department of Physics, Fudan University, Shanghai 200438, China}

\author{Cosimo~Bambi}
\email[Corresponding author: ]{bambi@fudan.edu.cn}
\affiliation{Center for Astronomy and Astrophysics, Department of Physics, Fudan University, Shanghai 200438, China}
\affiliation{School of Natural Sciences and Humanities, New Uzbekistan University, Tashkent 100000, Uzbekistan}

\author{Yi~Fan}
\affiliation{School of Physical Science and Technology, Soochow University, Suzhou 215006, China}
\affiliation{Center for Astronomy and Astrophysics, Department of Physics, Fudan University, Shanghai 200438, China}

\author{Temurbek~Mirzaev}
\affiliation{Center for Astronomy and Astrophysics, Department of Physics, Fudan University, Shanghai 200438, China}
\affiliation{Ulugh Beg Astronomical Institute, Uzbekistan Academy of Sciences, Tashkent 100052, Uzbekistan}
\affiliation{Institute for Advanced Studies, New Uzbekistan University, Tashkent 100007, Uzbekistan }

\author{Abdurakhmon~Nosirov}
\affiliation{Center for Astronomy and Astrophysics, Department of Physics, Fudan University, Shanghai 200438, China}
\affiliation{Institut f\"ur Astronomie und Astrophysik, Eberhard-Karls Universit\"at T\"ubingen, D-72076 T\"ubingen, Germany}

\author{Andrea~Santangelo}
\affiliation{Institut f\"ur Astronomie und Astrophysik, Eberhard-Karls Universit\"at T\"ubingen, D-72076 T\"ubingen, Germany}
\affiliation{Center for Astronomy and Astrophysics, Department of Physics, Fudan University, Shanghai 200438, China}

\begin{abstract}
Recently, we showed that the possibility of an interstellar mission to the closest black hole, while highly speculative and extremely challenging, is not completely unrealistic within the next few decades. Since such a mission might last around a century and require significant financial and human resources, it is crucial to assess whether it can truly study black holes and test General Relativity at levels unattainable by observational facilities in the Solar System for many years. In this manuscript, we assume the capability to decelerate the spacecraft and present a preliminary study of how probes orbiting a black hole could test the nature of the compact object.
\end{abstract}

\maketitle

%%%%%%%%%%%%%%%%% BODY OF PAPER %%%%%%%%%%%%%%%%%%

\section{Introduction}
\label{sec:intro} 

Proxima Centauri, the closest star to the Solar System, lies 4.24~light-years from the Sun. Clearly, exploring beyond the Solar System requires spacecraft that can travel at a significant fraction of the speed of light. In this regard, the Tsiolkovsky rocket equation unambiguously shows that our current chemically propelled rockets are completely unsuitable for such an ambitious goal. The Tsiolkovsky rocket equation reads $\Delta v = v_e \ln \left( m_i / m_f \right)$, where $\Delta v$ is the total change of the rocket's velocity, $v_e$ is the effective exhaust velocity of the propellant gases relative to the rocket, and $m_i$ and $m_f$ are the initial total mass (with propellant) and the final total mass (without propellant) of the rocket, respectively. Today, the most efficient chemical propellant is a combination of liquid hydrogen and liquid oxygen, with $v_e \sim 4.5$~km/s: even if $m_f$ were the proton mass and $m_i$ were the mass of the Earth, $\Delta v = 0.0018~c$, where $c$ is the speed of light, and our rocket would take over 2,000~years to reach Proxima Centauri. Thus, chemically propelled rockets are not suitable for exploration beyond the Solar System.

Studies on the development of spacecraft for interstellar missions began in the 1950s with the Orion project~\cite{Orion}, which aimed to reach Alpha Centauri in about 130~years using a large spacecraft accelerated by a series of controlled external nuclear explosions. In the 1970s, the British Interplanetary Society supported the Daedalus project~\cite{Daedalus}, a study of a fusion-powered spacecraft to reach Barnard's Star in about 50~years. In the 1980s, interest shifted from nuclear propulsion to beamed propulsion, originally proposed in the 1960s~\cite{Marx,Redding} and currently considered the most promising solution for the relatively near future (within a few decades from now). More speculative concepts have also been studied, ranging from antimatter rockets to the Alcubierre Warp Drive~\cite{Alcubierre:1994tu}.

Over the past 10-20~years, the exoplanet community has repeatedly discussed the possibility of interstellar missions to study exoplanets in nearby stellar systems~\cite{Lubin16,Lubin22,2018AcAau.152..370P,Kuhlmey25,Eubanks2026}. The most famous program was the Breakthrough Starshot Initiative,\footnote{\href{https://breakthroughinitiatives.org/}{https://breakthroughinitiatives.org/}} which aimed to develop a nanocraft capable of traveling at 1/5 the speed of light to reach Alpha Centauri in about 20~years but was abandoned in September~2025. Beamed propulsion circumvents the Tsiolkovsky rocket equation by avoiding the need to carry fuel onboard. The spacecraft (commonly called a nanocraft due to its very low weight) consists of a gram-scale wafer (constituting a fully functional probe with computer processor, navigation and communication equipment, etc.) and an extremely thin, meter-scale, dielectric metamaterial light sail. Ground-based or space-based high-power lasers hit the light sail, and the radiation pressure accelerates the nanocraft to its target velocity. There are no specific technical obstacles to reaching 90\% of the speed of light with this method, though higher velocities immediately increase mission costs.

In Refs.~\cite{Bambi:2025kcr,Bambi:2025hjn}, we discussed the possibility of sending one of these nanocrafts to the closest black hole, with the goal of testing black holes and General Relativity at levels likely out of reach for astrophysical observations from Earth or the Solar System. Simple considerations suggest that a few black holes may exist within 50~light-years of the Solar System, even if their detection is very challenging~\cite{Murchikova:2025oio,Nosirov:2026fjo}. If we discover a black hole 20-25~light-years from the Solar System and send a nanocraft traveling at 1/3 the speed of light, the nanocraft could reach the black hole in 60-75~years, perform a number of scientific experiments to test the nature of the compact object and the physics of strong gravitational fields, and send all data back to Earth. The data would arrive after 20-25~years, so the total mission duration would be of the order of 80-100~years. If the distance between the Solar System and the black hole is smaller or larger, the mission duration is correspondingly shorter or longer. If the nanocraft can travel at a higher velocity (requiring either more advanced technology or a larger mission budget), the mission duration is shortened.

In this manuscript, we begin the study of what kind of measurements should be performed near the black hole to test the nature of the compact object and the physics of strong gravitational fields. Addressing this question is important to determine whether an interstellar mission to a black hole indeed has the potential to test General Relativity with a precision and accuracy out of reach for astrophysical facilities on Earth or in the Solar System, even in the future, and whether it is therefore meaningful to engage in an 80-100~year project. For a preliminary study, we consider a simplified scenario in which the black hole is non-rotating and in vacuum. In reality, the black hole would have non-zero spin angular momentum and would be accreting from the interstellar medium. The density of the medium is low but increases as we approach the black hole, and its impact on the motion and measurements of our probes should be properly assessed.

A clear issue, not addressed here, is how to decelerate these tiny spacecraft upon arrival at the target system. Although several proposals exist in the literature, none appear to be directly applicable to our scenario. A light sail could exploit radiation pressure from a luminous stellar companion in the target system for photogravitational braking~\cite{2017ApJ...835L..32H,2017AJ....154..115H}. Electric and magnetic sails, conversely, could exchange momentum with charged particles, such as stellar-wind protons or the ionized interstellar medium~\cite{2016AcAau.128...13P,2017JPhCo...1d5007G,2020AcAau.168..146L}. The ``Halo Drive'' employs strongly deflected, returning photons to exchange energy with a moving black hole; although originally conceived for acceleration, the same mechanism could in principle be reversed for braking~\cite{Kipping:2019nqx}. If the target has a companion, a reverse gravity assist offers another possibility: a stationary flyby changes the probe's direction while preserving its asymptotic speed~\cite{Gould:2022}, whereas a moving companion can absorb part of its orbital energy~\cite{Dyson:1963,Gould:2022}. Relativistic slingshots of test particles around binary black holes have also been investigated~\cite{Zhang:2020}, and a reverse black-hole slingshot has been conceptually proposed for braking spacecraft at about 1/10 the speed of light~\cite{Gould:2022}. Arrival at 1/3 the speed of light would likely require a compact companion on a relativistic orbit and a favourable encounter geometry. A successful encounter could leave the probe bound to the target system, after which a smaller manoeuvre could place it onto the desired science orbit. These possibilities, however, require targets with very luminous or compact companions, which may not be the case for our black hole, as over 90\% of the black holes in the Galaxy are expected to be isolated, without any companion~\cite{Olejak:2019pln}.

In this manuscript (Paper~I), we assume that these nanocrafts can decelerate and enter certain orbits; we thus study how {\it orbiting probes} can test the nature of the compact object. In a second manuscript (Paper~II)~\cite{Fan:2026ule}, we assume that these nanocrafts cannot decelerate and study how we can test General Relativity with {\it flyby probes}.

The manuscript is organized as follows. In Section~\ref{s-mission}, we briefly review the phases of a possible mission to a nearby black hole. In Section~\ref{s-1}, we discuss the case of a single probe orbiting the black hole in a circular orbit. In Section~\ref{s-2}, we consider the case of two probes. In Sections~\ref{s-tests} and \ref{s-photons}, we discuss how these probes can measure possible deviations from the predictions of General Relativity. In Section~\ref{s-horizon}, we study how the probes can test the existence of the black hole's event horizon. Summary and conclusions are reported in Section~\ref{s-conclusions}.

%%%%%%%%%%%%%%%%%%%%%%%%%%%%%%%%%%%%%%%%%%%%%%%%%% 

\begin{figure*}[t]
\centering
\includegraphics[width=0.75\textwidth]{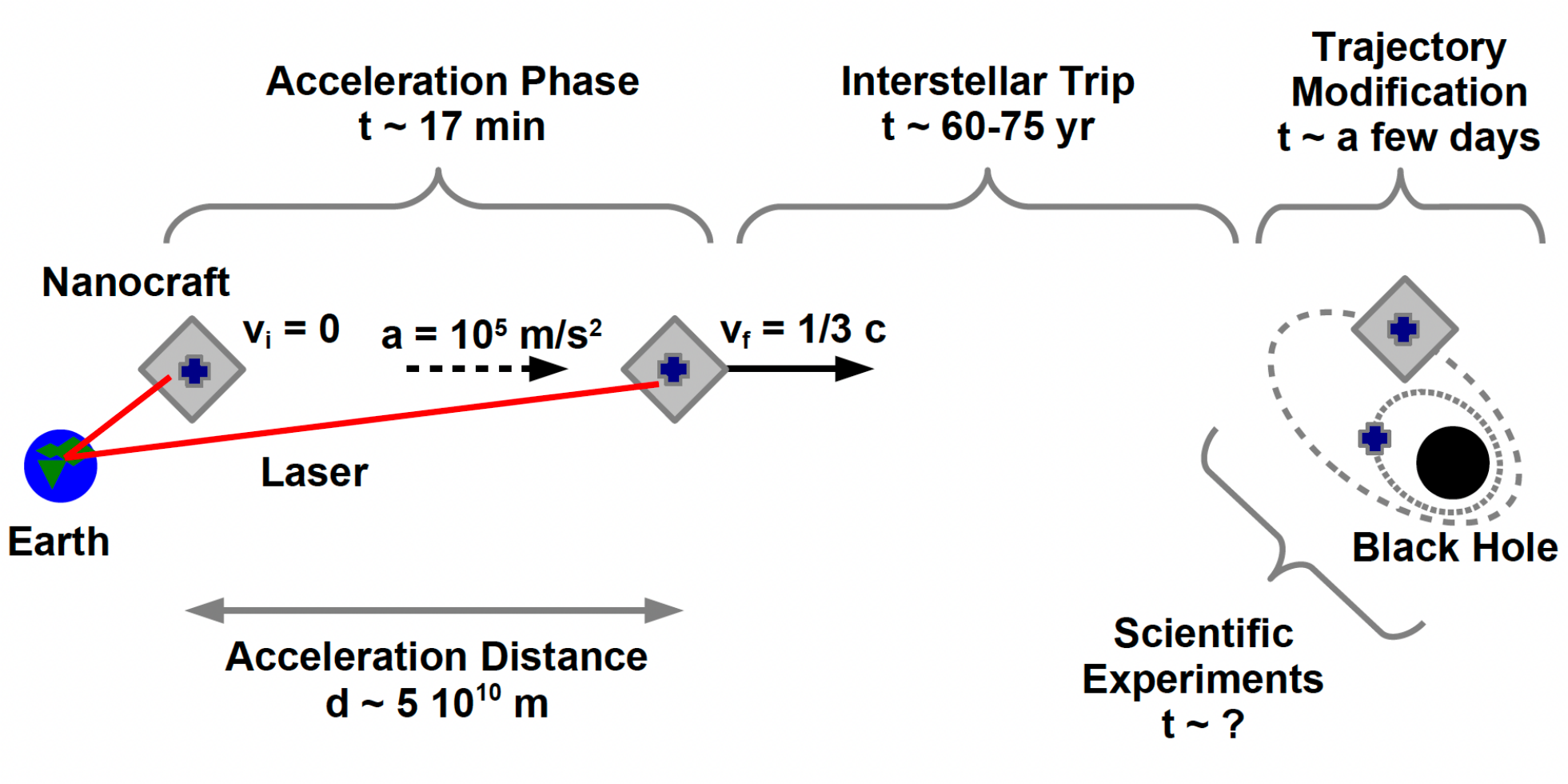}
\vspace{-0.2cm}
\caption{Phases of a hypothetical interstellar mission with a nanocraft to the closest black hole. See the text for details. Figure from Ref.~\cite{Bambi:2025kcr}. 
\label{f-m}}
\end{figure*}

\section{Interstellar mission to the closest black hole}\label{s-mission}

In this section, we briefly review the phases of the mission, without delving into details or discussing the challenges of each phase. The phases of such a mission are illustrated in the cartoon in Fig.~\ref{f-m}, and more details can be found in the original paper~\cite{Bambi:2025kcr} and references therein.

{\bf Phase~1: acceleration of the nanocraft.} The nanocraft must be accelerated from its initial velocity ($v_i \approx 0$) to its target velocity for the interstellar trip (for example, we can assume $v_f = c/3$, though this value is not important for the rest of the paper). High-power lasers strike the light sail to accelerate the nanocraft to its target velocity and along the correct direction to reach the black hole. The maximum acceleration of these nanocrafts is estimated to be $a \sim 10^5$~m~s$^{-2}$~\cite{Kuhlmey25}. If we accelerate the nanocraft from $v_i \approx 0$ to $v_f = c/3$ at the maximum acceleration, the acceleration phase lasts $t_1 = 1000$~s and at the end of this phase the nanocraft is at a distance $d \sim 5 \cdot 10^{10}$~m from Earth.

A rough estimate of the required laser power and total energy needed to accelerate the nanocraft can be made as follows. The initial momentum of the nanocraft is $p_i \approx 0$ and its momentum at the end of the acceleration phase should be $p_f = m v_f = m c / 3$, where $m$ is the mass of the nanocraft. If the frequency of the lasers is $\nu$, the momentum of each photon is $h\nu/c$ and we need $N \sim m c^2 / (3 h \nu)$ photons to accelerate the nanocraft. The total energy of these photons is $N h \nu \sim m c^2 / 3$. For $m = 3$~g, the total energy that must be delivered to the nanocraft is $E \sim 30$~GWh. Since this work should be done in $t_1 = 1000$~s, the required laser power is $P = E/t_1 \sim 100$~GW. When the nanocraft reaches its target velocity, the lasers are turned off, and the nanocraft starts its long journey to the target black hole.

{\bf Phase~2: interstellar trip.} Let us assume that the nanocraft travels at 1/3 of the speed of light. If the black hole is at 20-25~light years from the Solar System, the trip lasts 60-75~years. Minor corrections to the nanocraft's trajectory are possible using photon thrusters.

{\bf Phase~3: approach to the black hole and preparation for the scientific experiments.} When the nanocraft reaches the target black hole, it must prepare to carry out all the scientific experiments in the program. In this manuscript, we assume the nanocraft can decelerate and begin orbiting the black hole. In Paper~II~\cite{Fan:2026ule}, we show how flyby nanocrafts can test General Relativity and the spacetime geometry around the black hole without decelerating.

{\bf Phase~4: scientific experiments.} The nanocraft performs the scientific experiments in its program and sends the data to Earth. In Fig.~\ref{f-m}, we see two probes orbiting the black hole. A mothership orbits at larger radii, and a smaller probe orbits closer to the black hole. The mothership carries the nanocraft's light sail (the gray diamond in Fig.~\ref{f-m}), which can serve as an antenna to transmit all data back to Earth. The mothership cannot venture too close to the black hole because tidal forces from the compact object's gravitational field could destroy the light sail. The smaller probe, however, does not need an antenna to communicate with the mothership and can potentially approach very close to the black hole without being destroyed.

In the next sections, we will assume that we can decelerate our probes and place them into orbits around the target black hole. We aim to address the following question: what kind of measurements can these probes perform to test the nature of the compact object? The answer to this question is necessary to determine the requirements for the instruments onboard the probes (to be investigated in future studies) and the requirements on the probes' ability to enter specific orbits. Since this work is a preliminary study to address this question, we consider a simple framework: the black hole is non-rotating and in vacuum.

%%%%%%%%%%%%%%%%%%%%%%%%%%%%%%%%%%%%%%%%%%%%%%%%%%

\section{Case 1: One orbiting probe}\label{s-1}

\begin{figure}[t]
\centering
\includegraphics[width=0.4\textwidth]{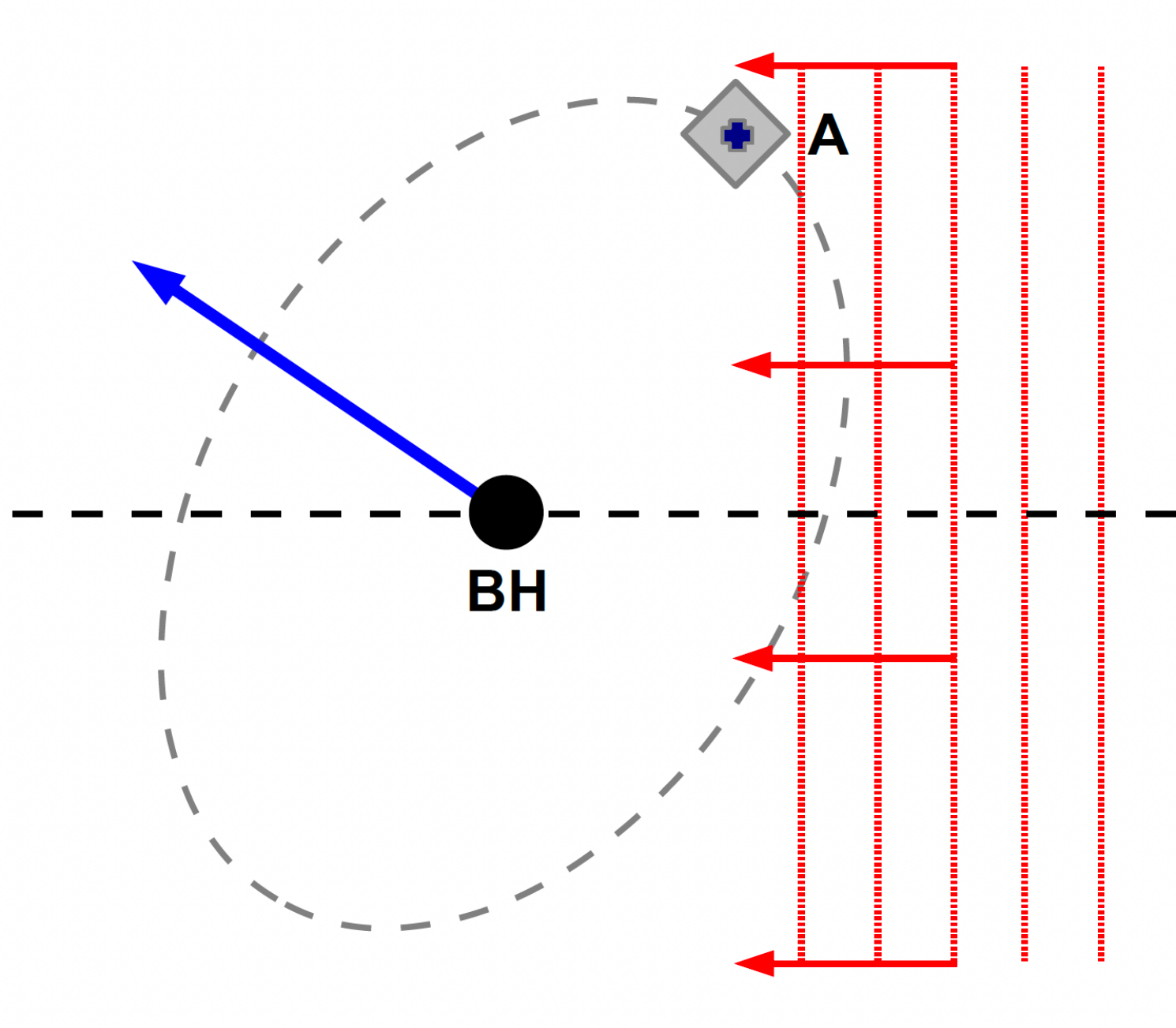}
\caption{Probe~A (A) is in a circular orbit around the black hole (BH). An electromagnetic signal sent from a station on Earth (red vertical lines) is used by probe~A to determine when it has completed an orbit around the black hole and to measure the ratio between its proper time and the time on Earth. In general, the angular momentum of probe~A (blue arrow) and the axis connecting the black hole to Earth (black-dashed line) are not aligned.
\label{f-1a}}
\end{figure}

In this section, we consider a single probe (probe~A) orbiting the black hole on a circular orbit, as illustrated in Fig.~\ref{f-1a}. We aim to determine what can be measured with a single probe.

In the Schwarzschild coordinates $\left( t , r , \theta , \phi \right)$, the line element of the Schwarzschild spacetime reads
\be\label{eq-ds2}
ds^2 &=& - \left( 1 - \frac{2 M}{r} \right) dt^2 
+ \left( 1 - \frac{2 M}{r} \right)^{-1} dr^2 \nonumber\\ && + r^2 d\theta^2 + r^2 \sin^2\theta d\phi^2 \, ,
\ee
where $M$ is the black hole mass parameter. Here and in the rest of the manuscript, we employ natural units in which $G_{\rm N} = c = 1$ and a metric signature $\left( - + + + \right)$.

From a station on Earth, we can send an electromagnetic signal toward the black hole, as shown in Fig.~\ref{f-1a} (the red vertical lines represent the wavefronts of the electromagnetic signal and the red horizontal lines with arrows indicate its propagation direction). This electromagnetic signal, detected by probe~A, serves two functions: $i)$ it allows probe~A to locate the position of Earth in the sky and thus to count the number of its orbits around the black hole; $ii)$ it enables probe~A, which must carry an onboard clock, to compare its proper time with the time on Earth.

We choose a coordinate system in which the orbit of probe~A lies on the equatorial plane of the Schwarzschild coordinates. The angular velocity of probe~A in the Schwarzschild coordinates is~\cite{Bambi:2017khi}
\be
\Omega = \frac{d\phi}{dt} = \frac{M^{1/2}}{r^{3/2}_{\rm A}} \, ,
\ee 
where $r_{\rm A}$ is the radial coordinate of probe~A. The relationship between a time interval measured in the proper time of probe~A, $\tau_{\rm A}$, and the same interval in the coordinate time $t$ is
\be \label{eqn:proper_coordinate_time_relation}
\Delta \tau_{\rm A} = \left( - g_{tt} - \Omega^2 g_{\phi\phi} \right)^{1/2} \Delta t 
= \left( 1 - \frac{3M}{r_{\rm A}} \right)^{1/2} \Delta t \, . \quad
\ee
$\Delta\tau_{\rm A} \rar 0$ as $r_{\rm A} \rar 3M$ because $r_\gamma = 3M$ is the radius of the photon orbit; the energy of a massive particle in a circular orbit diverges at the photon orbit, and no circular orbits exist for radial coordinates $r < r_\gamma$~\cite{Bambi:2017khi}. Circular orbits are stable for $r \ge 6 M$ and radially unstable for $3M \le r < 6M$; therefore, in what follows we will consider only orbits with radial coordinates $r \ge 6 M$.

We note that the time coordinate $t$ does not correspond to the time on Earth. Our proper time on Earth $\tau_{\rm Earth}$ is related to $t$ by (assuming no relative motion between the black hole and the Sun)
\be\label{eq-tEarth}
\Delta \tau_{\rm Earth} &\approx& \left( 1 - \frac{3}{2} \frac{M_\odot}{r_{\rm Earth}} 
- \frac{M_\oplus}{R_\oplus} \right) \Delta t \nonumber\\
&\approx& \left( 1 - 1.5 \cdot 10^{-8} \right) \Delta t \, , \hspace{0.5cm}
\ee
where $M_\odot$ is the mass of the Sun, $r_{\rm Earth}$ is the radius of Earth's orbit, $M_\oplus$ is the mass of the Earth, and $R_\oplus$ is the Earth's radius. After 3~days, the difference between $\Delta \tau_{\rm Earth}$ and $\Delta t$ is 4~ms, which becomes 40~ms after a month and 0.4~s after 10~months. As we will show below, such a difference must be taken into account -- with greater precision than the digits shown in Eq.~(\ref{eq-tEarth}) -- given the very high precision of the probe's measurements.

With the onboard clock and the detection of the electromagnetic signal from Earth, probe~A can measure the ratio between its proper time and the time coordinate $t$, i.e., $g_{\rm A} = \Delta\tau_{\rm A}/\Delta t$, as well as its angular velocity in its proper time, $\Omega_{\rm A}$, which is related to the Schwarzschild coordinates by
\be \label{eqn:Omega_A_proper_time}
\Omega_{\rm A} = \frac{d\phi}{dt} \frac{dt}{d\tau_{\rm A}} 
= \frac{M^{1/2}}{r^{3/2}_{\rm A}} \left( 1 - \frac{3M}{r_{\rm A}} \right)^{-1/2} \, .
\ee
From the measurement of $g_{\rm A}$ and $\Omega_{\rm A}$, probe~A can infer the black hole mass parameter $M$ and its radial coordinate $r_{\rm A}$
\be
M &=& \frac{1}{g_{\rm A} \, \Omega_{\rm A}} \left( \frac{1 - g^2_{\rm A}}{3} \right)^{3/2} \, , \\
r_{\rm A} &=& \frac{1}{g_{\rm A} \, \Omega_{\rm A}} \left( \frac{1 - g^2_{\rm A}}{3} \right)^{1/2} \, .
\ee
Probe~A can count its orbits around the black hole by detecting the electromagnetic signal from Earth. It must also carry an onboard clock, which would allow it to measure $\Omega_{\rm A}$. Using a gyroscope or observing reference stars may require instruments that are too heavy for the interstellar mission.

The precision of the measurements of $M$ and $r_{\rm A}$ depends on probe~A's ability to measure $\Omega_{\rm A}$ and $g_{\rm A}$. If probe~A can complete many orbits around the black hole, obtaining very precise measurements of $\Omega_{\rm A}$ and $g_{\rm A}$ is relatively straightforward. For example, if the clock onboard probe~A has a precision of 1~$\mu$s and probe~A orbits the black hole for 3~days (with respect to the coordinate time $t$), we find\footnote{The relative uncertainty on the measurement of $\Omega_{\rm A}$ is $\Delta \Omega_{\rm A}/\Omega_{\rm A} = \Delta T/T$, where $\Delta T$ is the uncertainty on the orbital period of probe~A and $T$ is the orbital period. $\Delta T$ is given by the ratio of the clock sensitivity to the number of orbits. Since $T$ equals the ratio of the experiment duration to the number of orbits in that interval, $\Delta T/T$ reduces to the ratio of the clock sensitivity to the experiment duration, and is thus independent of both the number of orbits and the orbital radius of the probe.} 
\be
\frac{\Delta \Omega_{\rm A}}{\Omega_{\rm A}} \sim 4 \cdot 10^{-12} \, ,
\ee
which becomes $4 \cdot 10^{-13}$ for 1~month and $4 \cdot 10^{-14}$ for 10~months. This result is independent of the orbital radius of the probe. Achieving such a precise measurement of $\Omega_{\rm A}$ requires that we account for Eq.~(\ref{eq-tEarth}).

The uncertainty on $g_{\rm A}$ is determined by the uncertainties on $\Delta\tau_{\rm A}$ and $ \Delta t$, though the latter is  presumably negligible compared to the former. The relative uncertainty on $\Delta\tau_{\rm A}$ should be of order the ratio of the clock sensitivity (1~$\mu$s in the example above) to the experiment duration measured in the probe's proper time. The latter is comparable to $ \Delta t$. For $r_{\rm A} = 6M$, we have $g_{\rm A} = 0.71$; as $r_{\rm A}$ increases, $g_{\rm A} \rar 1$. For $r_{\rm A} = 100M$, we already have $g_{\rm A} = 0.9849$. Therefore, $\Delta g_{\rm A}/g_{\rm A}$ is similar to $\Delta \Omega_{\rm A}/\Omega_{\rm A}$. We thus find that we can measure $M$ and $r_{\rm A}$ with very high precision, comparable to the precision of the measurements of $\Omega_{\rm A}$ and $g_{\rm A}$, at least within our simplified framework.

Note that assuming a constant clock error is a very crude estimate; a real oscillator contains a spectrum of error processes that depend on the measurement timescale. For a nanocraft whose primary requirement is precise short-term interval timing -- rather than absolute time after decades -- the relevant error sources shift. Deterministic, controllable errors such as the initial frequency offset and the long-term aging rate can be calibrated and partially compensated by a low-order polynomial model (e.g., a linear frequency drift term). However, the extreme deep-space environment makes environmentally induced frequency fluctuations a dominant and largely unpredictable source of error: residual temperature variations, magnetic field changes, and power-supply ripple all directly perturb the oscillator's frequency. If the nanocraft maintains intermittent communication with Earth, the accumulating frequency drift can be measured on the ground and corrected via occasional updates, effectively resetting the long-term error growth. Consequently, an optimal measurement timescale exists -- a middle ground where short-term stochastic phase noise (white and flicker phase) has been averaged to a low level, while the accumulated error from residual environmental drift and aging has not yet become severe.

If the angular momentum of probe~A is parallel or antiparallel to the axis connecting the black hole to Earth, probe~A detects an electromagnetic signal that is constant in time. In general, this is not the case (as shown in Fig.~\ref{f-1a}), and the orientation of probe~A's angular momentum relative to the Earth-black hole axis can be described by two angles, say $\left( \alpha , \beta \right)$. The electromagnetic signal will be periodically redshifted and blueshifted during each orbit, and from the temporal evolution of the detected signal it is possible to infer $\left( \alpha , \beta \right)$.

At this stage, we have inferred $M$, $r_{\rm A}$, and possibly even $\left( \alpha , \beta \right)$, assuming the Schwarzschild metric. To test whether the spacetime around the black hole is descried by the Schwarzschild solution, additional measurements are required. If probe~A can move to another circular orbit, it can repeat the measurements to infer the black hole mass $M$ and its new radial coordinate $r_{\rm A}'$. If the central object is not a Schwarzschild black hole, the two measurements of the black hole mass $M$ may differ. Another option, which will be discussed in Section~\ref{s-photons}, is to use highly collimated electromagnetic signals that can orbit the black hole and return to the probe.

%%%%%%%%%%%%%%%%%%%%%%%%%%%%%%%%%%%%%%%%%%%%%%%%%%

\begin{figure}[t]
\centering
\includegraphics[width=0.4\textwidth]{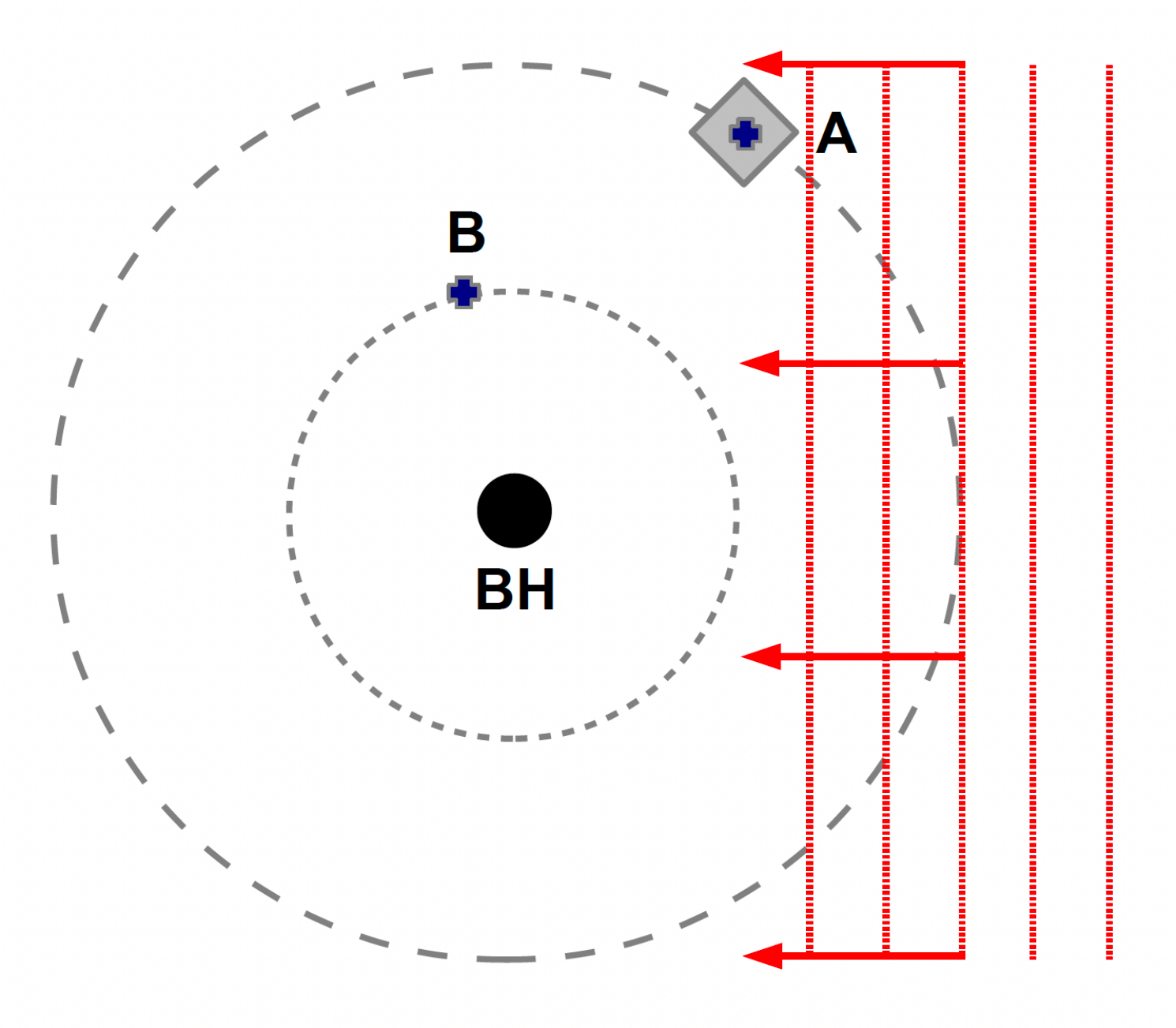}
\caption{As in Fig.~\ref{f-1a} with two probes: probe~A and probe~B. Probe~A is the mothership with the light sail (the gray diamond in the cartoon), which can be used as an antenna to send all data to Earth. Probe~A cannot orbit too close to the black hole. Probe~B does not need a large antenna to communicate with probe~A, and therefore it can orbit close to the black hole.
\label{f-2a}}
\end{figure}

\section{Case 2: Two orbiting probes}\label{s-2}

Let us now assume we have two probes orbiting the black hole, as shown in Fig.~\ref{f-2a}. Probe~A can be thought of as the mothership: it carries the light sail, which can be used as an antenna to transmit all data to Earth, and it cannot orbit too close to the black hole, otherwise tidal forces could easily destroy the light sail. Probe~B is a smaller probe: it does not need a large antenna to communicate with probe~A and can therefore orbit closer to the black hole without being destroyed by tidal forces near the compact object.

If probe~B can detect the electromagnetic signal from the station on Earth and has an onboard clock, it can measure the ratio between its proper time and the time coordinate $t$, $g_{\rm B}$, as well as its angular velocity in its proper time, $\Omega_{\rm B}$. Just like probe~A, probe~B can infer the black hole mass $M$ and its radial coordinate $r_{\rm B}$:
\be
M &=& \frac{1}{g_{\rm B} \, \Omega_{\rm B}} \left( \frac{1 - g^2_{\rm B}}{3} \right)^{3/2} \, , \\
r_{\rm B} &=& \frac{1}{g_{\rm B} \, \Omega_{\rm B}} \left( \frac{1 - g^2_{\rm B}}{3} \right)^{1/2} \, .
\ee
With two probes, we obtain two independent measurements of the black hole mass $M$. The two measurements should be consistent if the spacetime geometry around the black hole is described by the Schwarzschild solution, but they are likely to be inconsistent if the compact object is not a Schwarzschild black hole. As shown in the previous section, a probe can measure $M$ with very high precision. With two probes, we can perform a consistency test of the Schwarzschild spacetime by comparing the mass measurements from the two probes. More stringent tests can be obtained if probe~B orbits as close as possible to the black hole (while still satisfying $r_{\rm B} \ge 6M$), though this depends on probe~B's ability to maneuver into such orbits with very small radial coordinates.

%%%%%%%%%%%%%%%%%%%%%%%%%%%%%%%%%%%%%%%%%%%%%%%%%%

\section{Constraining deviations from the Schwarzschild metric}\label{s-tests}

In the previous sections, we have assumed that the spacetime geometry around the black hole is described by the Schwarzschild solution and we have discussed possible consistency tests. In this section, we consider a deformed Schwarzschild metric and we study the ability of the two probes to constrain the deformation parameter of the spacetime. This study can illustrate better the constraining power of the two orbiting probes with respect to present and future astrophysical observations~\cite{Bambi:2015kza,Yagi:2016jml}.

We consider the Johannsen spacetime with the deformation parameter $\alpha_{13}$~\cite{Johannsen:2013szh}. If $\alpha_{13} = 0$, we recover the black holes of General Relativity (i.e., Schwarzschild black holes in the non-rotating case and Kerr black holes in the presence of a non-vanishing spin angular momentum) while any non-vanishing value of the deformation parameter $\alpha_{13}$ would describe a deviations from the predictions of Einstein's gravity. Current observational constraints on the parameter $\alpha_{13}$ are summarized in Ref.~\cite{Das:2026zyt} (see Ref.~\cite{Bambi:2022dtw} for more details). From X-ray observations, the most precise and accurate measurement of this deformation parameter is $\alpha_{13} = -0.02_{-0.14}^{+0.03}$ (3-$\sigma$), which is obtained from the black hole binary GX~339--4~\cite{Tripathi:2020dni,Tripathi:2020yts}. From gravitational wave observations, the most stringent constraint is $\alpha_{13} = 0.07_{-0.16}^{+0.16}$ (3-$\sigma$) from event GW230627\_15337~\cite{Das:2026zyt}. The future Laser Interferometer Space Antenna (LISA), which is currently expected to operate from 2035 to 2045, will observe gravitational waves from extreme mass ratio inspirals (EMRIs), where a stellar-mass compact object (stellar-mass black hole, neutron star, or white dwarf) orbits a supermassive black hole of $\sim 10^6~M_\odot$. Since LISA can observe the gravitational wave signal from such systems for millions of orbits, we expect to be able to measure the deformation parameter $\alpha_{13}$ with a precision of 0.01, and possibly even better for some systems~\cite{Barack:2006pq}.

Since our goal is to test General Relativity, in this preliminary work we focus on the case of non-rotating black holes and examine how deviations from the Schwarzschild solution could be detected. The rotating case is significantly more complicated, not only because it introduces an extra parameter (the spin), but also because the probe's motion then depends on the angle between the black hole spin axis and the probe's angular momentum -- whereas here we assume, without loss of generality, that the motion lies in the equatorial plane. In reality, the black hole spin can be expected to have a stronger effect on the probe's motion than possible deviations from the Schwarzschild solution: in a multipole expansion, the spin appears as a dipole term, while any deviation from General Relativity is expected to enter at the quadrupole or higher order. We also note that the Johannsen spacetime with deformation parameter $\alpha_{13}$ does not encompass all physically relevant static and spherically symmetric deviations from the Schwarzschild geometry. Our test of the Schwarzschild metric is therefore model-dependent. Nevertheless, as shown in Ref.~\cite{2024PhRvD.109f4059Z}, this kind of test remains a valid tool for discovering new physics, even though its constraining power may be lower than that of a test employing the correct black hole metric.

For a vanishing spin angular momentum, the line element of the Johannsen spacetime with the deformation parameter $\alpha_{13}$ reads 
\be\label{eq-j}
ds^2 &=& - \left( 1 - \frac{2 M}{r} \right) 
\left( 1 + \alpha_{13} \frac{M^3}{r^3} \right)^{-2} dt^2 \nonumber\\ 
&& + \left( 1 - \frac{2 M}{r} \right)^{-1} dr^2 
+ r^2 d\theta^2 + r^2 \sin^2\theta d\phi^2 \, , \qquad
\ee
which reduces to the Schwarzschild metric for $\alpha_{13} = 0$. Continuing with the setup with two probes in circular orbits, the angular velocity $\Omega_{\rm A} = d\phi/d\tau_{\rm A}$ and the redshift factor $g_{\rm A}=\Delta\tau_{\rm A}/\Delta t$ can be expressed now as
\be
\Omega_{\rm A} = \sqrt{ 
    \frac{ M r_{\rm A}^3 \bigl[ r_{\rm A}^3 + \alpha_{13} M^2 (3r_{\rm A} - 5M) \bigr] }
         { \alpha_{13} M^3 (3M - 2r_{\rm A}) r_{\rm A}^5 + r_{\rm A}^8 (r_{\rm A} - 3M) }
}, \\
g_{\rm A} = \sqrt{ 
    \frac{ \alpha_{13} M^3 (3M - 2r_{\rm A}) r_{\rm A}^5 + r_{\rm A}^8 (r_{\rm A} - 3M) }
         { (\alpha_{13} M^3 + r_{\rm A}^3)^3 }
}. 
\ee
The quantities $\Omega_{\rm B}$ and $g_{\rm B}$ can be expressed similarly in terms of $M$, $\alpha_{13}$, and $r_{\rm B}$. These four measurements depend strictly on $M$, $\alpha_{13}$, $r_{\rm A}$, and $r_{\rm B}$. Although we cannot isolate $\alpha_{13}$ algebraically into a simple formula, we can easily solve this system numerically to infer its value.

Let us now consider probes moving in eccentric orbits. When comparing orbital dynamics between the Schwarzschild and Johannsen metrics, one must be cautious about whether the results obtained from two distinct metrics are truly comparable, as the definitions and physical meanings of coordinates and physical quantities can differ significantly.  
A major challenge lies in generating consistent initial conditions. It is physically impossible to construct an initial state in both metrics that simultaneously shares the same specific energy $E$, specific angular momentum $L_z$, initial radius $r$, and initial radial velocity $dr/d\tau$ while remaining consistent with the geodesic equations of motion in General Relativity.

Therefore, to make a mathematically and physically rigorous comparison, we establish the initial conditions of the probes as follows:
\begin{itemize}
  \item First, we select an eccentric orbit with periapsis ($r_p$) and apoapsis ($r_a$) in the Schwarzschild metric, characterized by specific energy $E$ and specific angular momentum $L_z$. 
  \item Next, we require the probes in the Johannsen metric to share this exact same coordinate periapsis and apoapsis. The simulations for both spacetimes are initialized at apoapsis, where $dr/d\tau = 0$.
  \item Finally, the specific energy and angular momentum for the probe in the Johannsen metric are derived directly from these turning points. Consequently, the assigned $E$ and $L_z$ values will differ between the two spacetimes, though the discrepancy stays negligibly small for $|\alpha_{13}| \ll 1$. 
\end{itemize}
Although we sacrifice the equality of $E$ and $L_z$ between the two spacetimes, this setup offers several critical advantages.  First, because the spatial components of the Johannsen metric in Eq. \eqref{eq-j} remain undeformed, the radial coordinate $r$ retains the exact same geometric definition -- the areal radius -- as in the Schwarzschild metric. Consequently, matching the coordinate periapsis $r_p$ and apoapsis $r_a$ across both spacetimes physically constrains the orbits to share identical radial turning surfaces, giving the comparison strict physical meaning. Second, the constants of motion $E$ and $L_z$ are strictly dependent on the underlying metric of the spacetime. Consequently, their numerical values are not physically comparable across different geometries, even if we artificially set them to be equal. Third, by fixing the periapsis and apoapsis, we establish a clean baseline to compare universally well-defined physical observables, such as the radial period and apsidal precession per orbit.

A useful quantity to distinguish a Schwarzschild from a deformed-Schwarzschild spacetime is the apsidal precession rate. We define the precession rate $\Delta\Phi$ as the excess azimuthal angle swept out between consecutive radial turning points per revolution:
\be
  \Delta\Phi = \Delta\phi(T_r)-2\pi,
\ee
where $\Delta\phi(T_r)$ is the total change in the $\phi$ coordinate over one radial period $T_r$, defined as the proper time elapsed from one periapsis (or apoapsis) to the next. 
In Table~\ref{tab:comparison_precession}, we catalog several orbital configurations across varying eccentricities $e$, periapsis $r_p$, and deformation parameter $\alpha_{13}$. Finally, we calculate the observational timescale $\tau_{\text{obs}}$ in terms of the proper time of the probe, representing the approximate physical time required to accumulate a measurable anomalous phase shift of $\delta\phi = \pi$ due to the difference in the precession rate between two metrics. Note that $\delta \phi = n \Delta \Phi$ and $\tau_{\text{obs}} = n T_r$, where $n$ is the number of orbits (or radial periods $T_r$) required to accumulate the phase shift $\delta \phi$.
There are several core takeaways from this data:
\begin{itemize}
  \item \textbf{Proximity to the black hole:} Closer orbits significantly increase the chances of detecting a deformation effect in a shorter amount of time. If a probe is sent extremely close to the innermost stabler circular orbit ($r_{\text{ISCO}}=6M$), such as the example orbit with $r_p=10.0M$ and $e=0.1$, the spacetime deformation becomes overwhelmingly apparent.
  \item \textbf{The role of eccentricity:} Increasing orbital eccentricity does not uniformly enhance detectability. As demonstrated in Table~\ref{tab:comparison_precession}, increasing eccentricity while holding the periapsis $r_p$ constant actually results in an increase in $\tau_{\text{obs}}$. This effect arises because a higher eccentricity at a fixed $r_p$ corresponds to a larger apoapsis $r_a$, thereby shifting the time-averaged orbital radius further from the black hole. Conversely, if a highly eccentric orbit drives the periapsis deeper into the strong-field regime, detectability improves significantly.
  \item \textbf{The dependency on $\alpha_{13}$:} For fixed $r_p$ and $e$, $\tau_{\text{obs}}$ scales almost linearly with $\alpha_{13}$. 
\end{itemize}

\begin{table}[t]
    \centering
    \caption{Observability timescale $\tau_{\text{obs}}$ required for an accumulated phase difference
             $\delta\phi = \pi$, for selected orbital parameters $r_p$ and $e$ and deformation parameter $\alpha_{13}$. The mass of the black hole is set to $10~M_\odot$ for obtaining the values for $\tau_{\text{obs}}$.}
    \label{tab:comparison_precession}
    \begin{tabular}{|c|c|c|c|c|}
        \hline
        \multicolumn{2}{|c|}{Orbit} &
        \multicolumn{3}{|c|}{$\tau_{\text{obs}}$ ($\delta\phi = \pi$)} \\
        \hline
        $r_p$ [M] & $e$ & $\alpha_{13} = 10^{-3}$ & $\alpha_{13} = 10^{-5}$ & $\alpha_{13} = 10^{-7}$ \\
        \hline
        10 & 0.1 & 0.0370 hours & 3.70 hours &  370 hours \\
        \hline
        10 & 0.3 & 0.0840 hours & 8.40 hours & 840 hours \\
        \hline
        100 & 0.1 & 196 hours & $1.96\times10^4$ hours& $1.96\times10^6$ hours\\
        \hline
        100 & 0.3 & 401 hours & $4.01\times10^4$ hours & $4.01\times10^6$ hours \\
        \hline
        1000 & 0.1 & 73.1 years & $7.31\times10^3$ years& $7.29\times10^5$ years\\
        \hline
        1000 & 0.3 & 149 years & $1.49\times10^4$ years& $1.47\times10^6$ years \\
        \hline
    \end{tabular}
\end{table}

%%%%%%%%%%%%%%%%%%%%%%%%%%%%%%%%%%%%%%%%%%%%%%%%%%

\section{Testing the spacetime with photons}\label{s-photons}

In this section, we consider the possibility of testing the Schwarzschild metric using highly collimated electromagnetic signals (photons) emitted by probes orbiting the black hole. These photons can orbit the black hole and then return to the probes. In the Schwarzschild metric, we can make very clear predictions regarding both the exact emission direction of the photon and the time it takes for the photon to return to the probe. Because the photon must orbit near the photon orbit when it approaches the black hole, and the photon orbit is unstable, even small deviations from the Schwarzschild metric can produce large differences in the photon's trajectory and in the return time to the probe.

\begin{figure*}[t]
\centering
\includegraphics[width=0.45\textwidth]{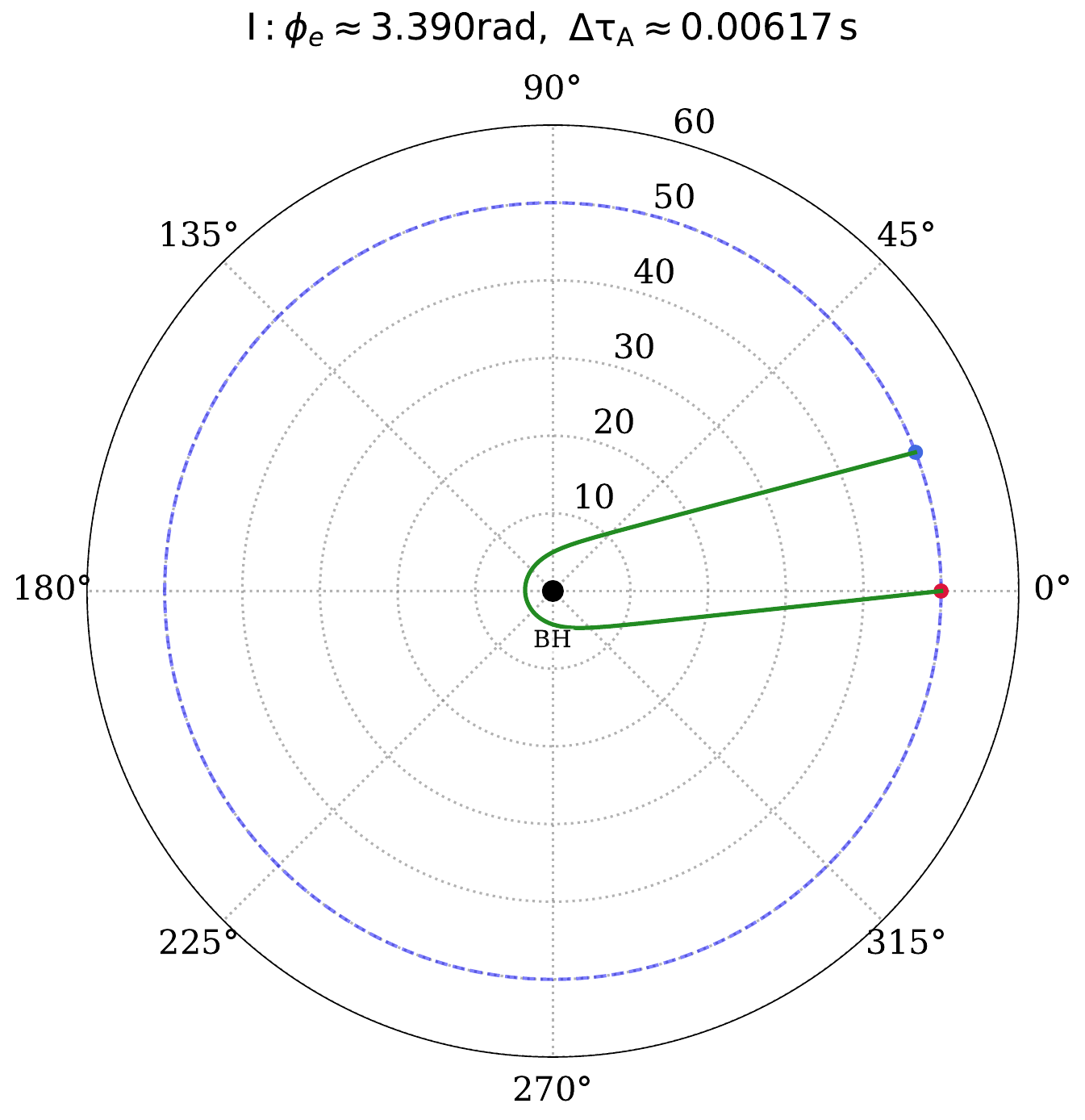}
\hspace{0.5cm}
\includegraphics[width=0.45\textwidth]{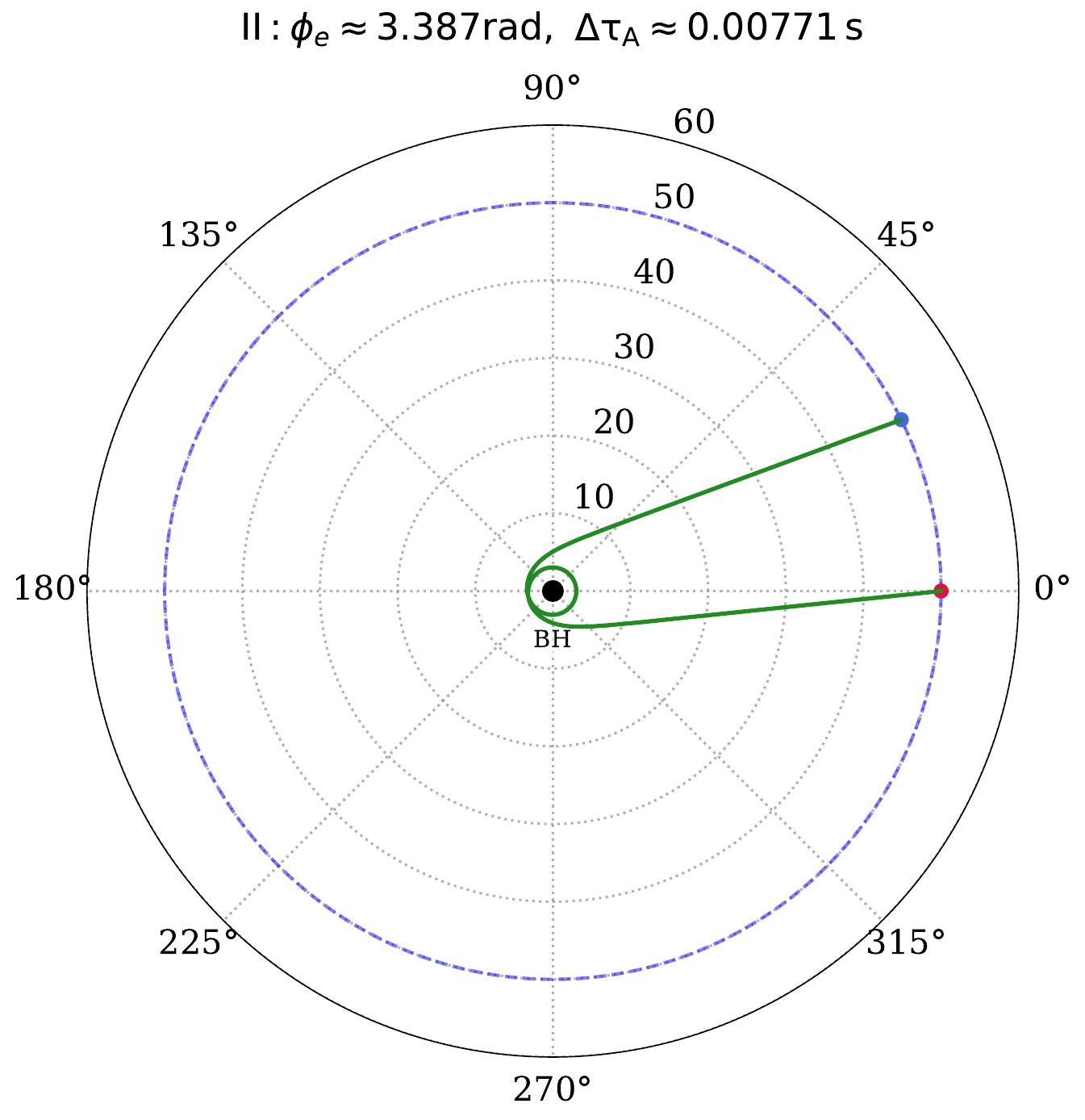}\\
\vspace{0.8cm}
\includegraphics[width=0.45\textwidth]{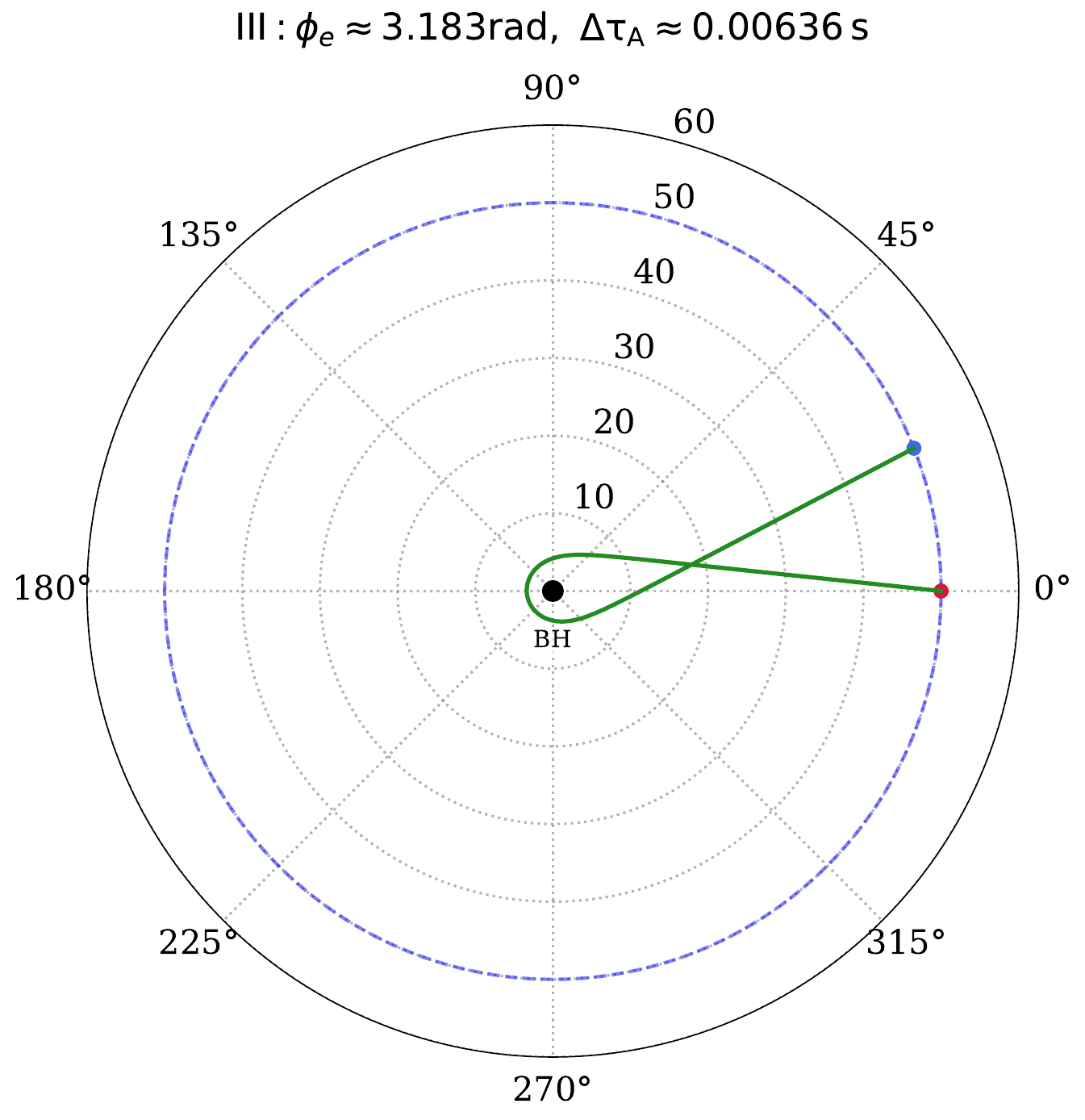}
\hspace{0.5cm}
\includegraphics[width=0.45\textwidth]{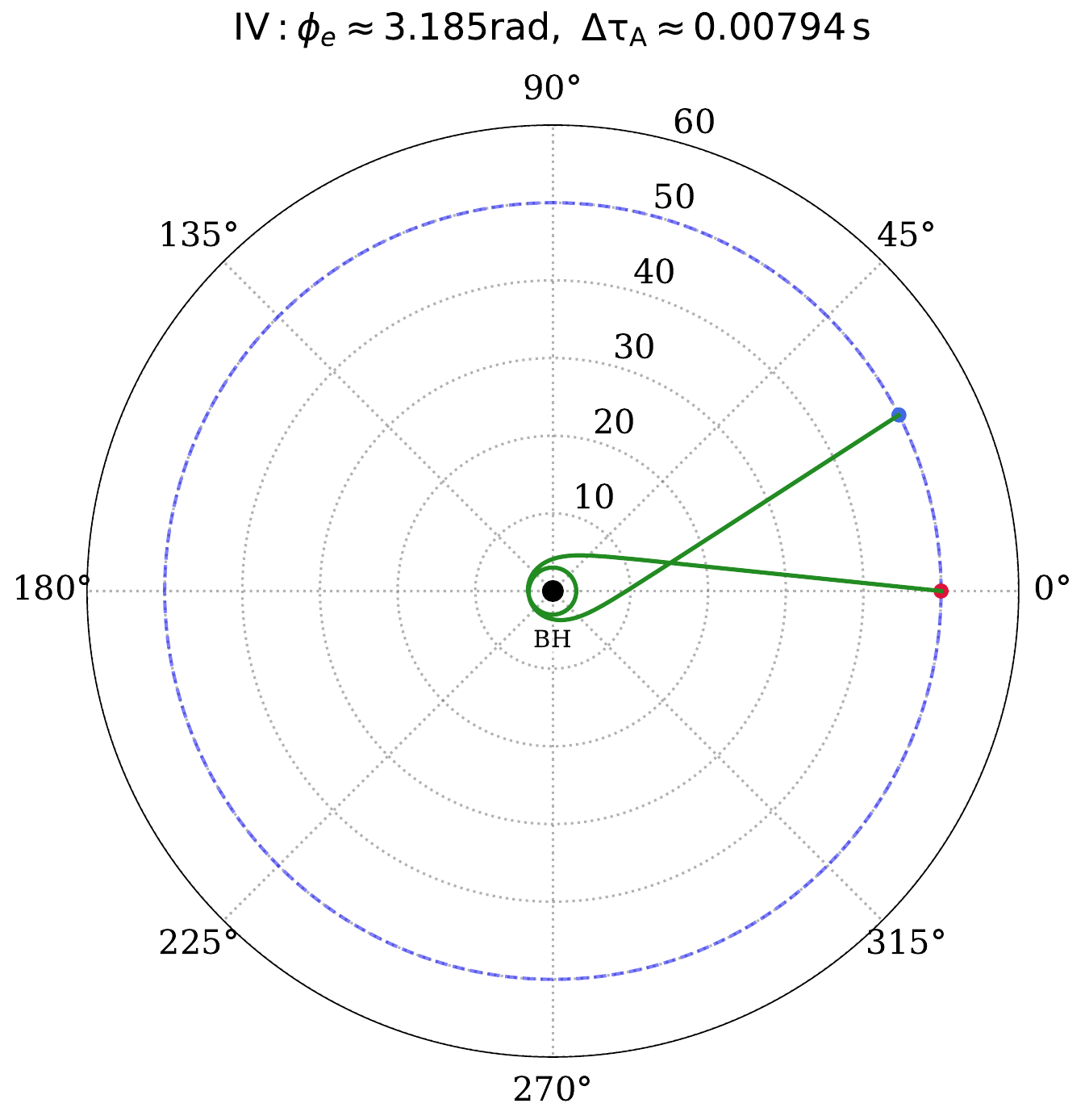}
\caption{For the one probe scenario, four different examples show that the emitted signals can be received by the probe itself after a time $\Delta \tau_{\rm A}$. In cases II and IV, the photon orbits near the photon sphere for one full circle before returning to the probe. In the caption of each plot, $\phi_{\rm e}$ is the emission angle in the local reference frame of the probe, and $\Delta \tau_{\rm A}$ is the travel time of the photon in the probe's proper time. }
\label{f-1c}
\end{figure*}

We begin with the case of one orbiting probe (probe~A). In the Schwarzschild metric, given the black hole mass $M$ and the orbital radius of probe~A $r_{\rm A}$, we can calculate the exact direction in which probe~A must send the signal, as well as the time interval from emission to reception, $\Delta \tau_{\rm A}$. This problem has more than one solution, as the signal can make one or more orbits around the black hole. In Fig.~\ref{f-1c}, we consider an orbital radius $r_{\rm A} = 50~M$ and show four trajectories with different orbits around the black hole. We assume $M = 10~M_\odot$ for estimating $\Delta \tau_{\rm A}$. In Appendix~\ref{app:local_frame}, we construct the locally Minkowskian reference frame for the orbiting probe. The emission angle $\phi_{\rm e}$ is reported in Fig.~\ref{fig:local_tetrad_spherical}, and we take $\theta_{\rm e}=\pi/2$ for our cases. In case~I, probe~A must emit the electromagnetic signal at an angle $\phi_{\rm e} \approx 3.390$~rad in the probe's local reference frame and will measure a time interval $\Delta \tau_{\rm A} \approx  0.00617$~s between emission and detection. In case~II, we have $\phi_{\rm e} \approx 3.387$~rad and $\Delta \tau_{\rm A} \approx 0.00771$~s. In case~III, we have $\phi_{\rm e} \approx 3.183$~rad and $\Delta \tau_{\rm A} \approx 0.00636$~s. In case~IV, we have $\phi_{\rm e} \approx 3.185$~rad and $\Delta \tau_{\rm A} \approx 0.00794$~s. In the case of a non-Schwarzschild metric around the black hole, the values of $\theta_{\rm A}$ and $\Delta \tau_{\rm A}$ change.

Let us now move to the case of two probes (probe~A and probe~B). Probe~B emits a photon toward the black hole; this photon orbits the compact object and is then received by probe~A. The condition for the photon emitted by probe~B to be received by probe~A is given by\footnote{We note that Eq.~(\ref{eqn:emission_receiving_condition}) requires that probe~A is in a circular orbit, while the motion of probe~B has no restrictions.}
\begin{equation}\label{eqn:emission_receiving_condition}
  \phi_{\rm B}(t_{\rm e})+\Delta \phi(\Delta t)-\left[ \phi_{\rm A}(t_{\rm e})+\Omega_{\rm A} \Delta t + 2n\pi \right] = 0 \, ,
\end{equation}
where $t_{\rm e}$ is the coordinate time $t$ at which probe~B emits the photon, $\phi_{\rm A}(t_{\rm e})$ and $\phi_{\rm B}(t_{\rm e})$ are the coordinate $\phi$ values of probe~A and probe~B at coordinate time $t_{\rm e}$, respectively, $\Delta t$ is the coordinate time interval from emission by probe~B to detection by probe~A, and $\Delta \phi(\Delta t)$ is the change in the photon's coordinate $\phi$ during the time interval $\Delta t$\footnote{$\Delta t$ and $\Delta \phi(\Delta t)$ depend on the radius $r$ and radial velocity $dr/d\tau$ of probe~B, as well as the local emission angle in probe~B's local reference frame. $\Delta t$ and $\Delta \phi(\Delta t)$ are computed through high-precision ray-tracing.}. $\Omega_{\rm A}$ is the angular velocity of probe~A. By solving Eq.~\eqref{eqn:emission_receiving_condition} with different integer values of $n$, we can find a local emission angle of the photon for each $n$ and each position of probe~B. If $\phi_{\rm A}$ and $\phi_{\rm B}$ are close to each other, the integer $n$ approximately represents the number of circles the photon makes around the photon sphere before being received by probe~A. $n=0$ corresponds to direct emission. $n>0$ indicates that the photon orbits counterclockwise, and $n<0$ indicates clockwise orbits.

We consider an example to demonstrate the possibility of testing the black hole spacetime with this approach. We assume that we can measure the energy, arrival time, and incident angle of multiple electromagnetic signal echoes (e.g., $n = -2$, $-1$, 0, 1, and 2) emitted by a probe. Using this incident information, we can formulate a non-linear optimization problem via backward ray-tracing. By treating the affine parameters ($\lambda_i$) of each corresponding null geodesic as free variables, we force the system to converge at a single, shared 4D emission event. This geometric requirement yields a heavily overdetermined system, allowing us to simultaneously fit for the background spacetime parameters, such as the black hole mass $M$ and the deformation parameter $\alpha_{13}$. Note that the trajectory information of probe~A and probe~B are not necessary to be determined. Once the exact emission coordinate is localized, the redshift data uniquely determine the local 4-velocity of the emitting source. If we can pre-measure the trajectories of probe~A and probe~B using the technique described in Section~\ref{s-1}, even stronger constraints on the black hole mass and the deformation parameter can be obtained. Ultimately, this approach establishes a stringent, self-consistent null test for General Relativity: any fundamental theoretical deviation from the true spacetime geometry will inherently break the geometric convergence of the backward-traced geodesics, preventing a zero-residual vertex solution.

\begin{figure}[t]
\centering
\includegraphics[width=0.48\textwidth]{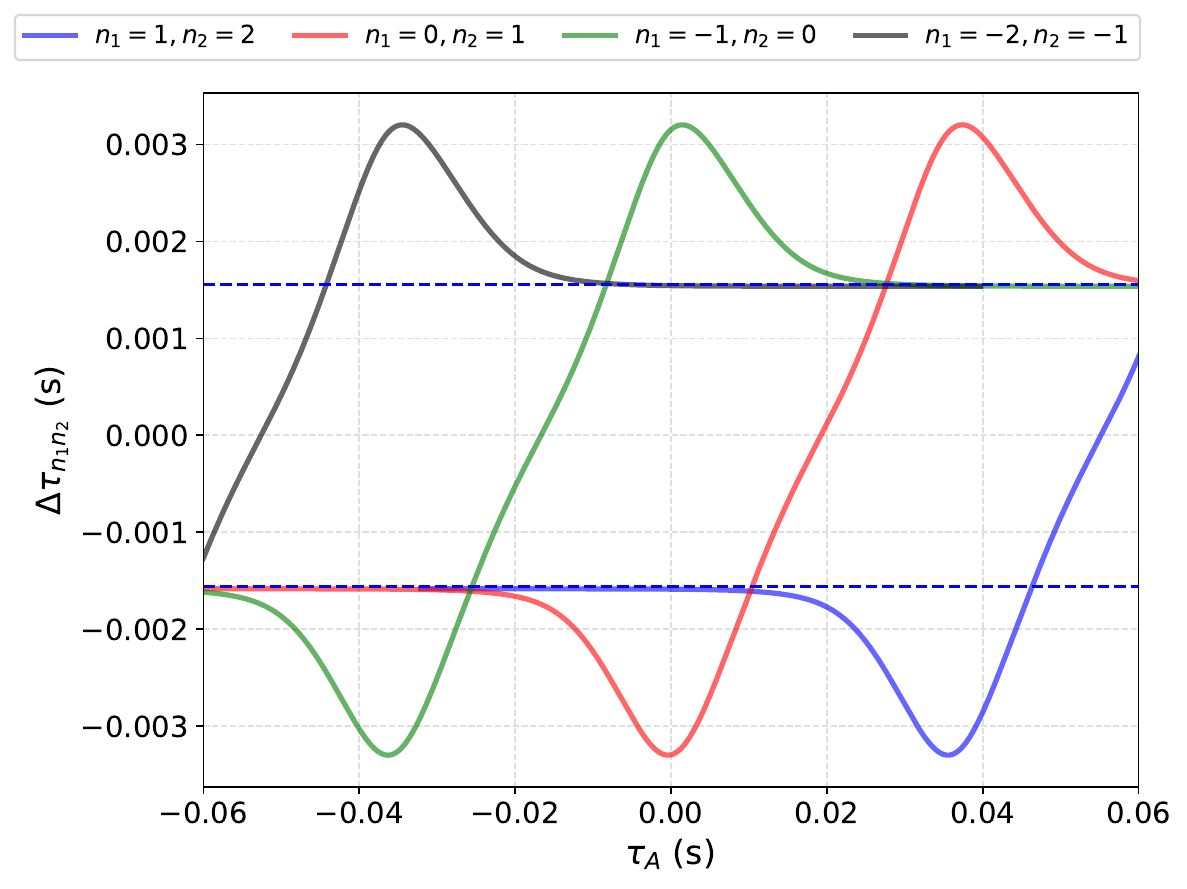}
\vspace{-0.6cm}
\caption{Time difference $\Delta\tau_{n_1n_2}$ (measured by probe~A) between the arrival times of photons with adjacent orders of $n$ as a function of the photon emission time $\tau_{A}$. Every curve asymptotes to $\pm T_\gamma (1 - 3M/r_{\rm A})^{-1/2}$ (blue dashed lines), where $T_\gamma$ of the orbital period at the photon orbit $r = r_\gamma$.}
\label{fig:time_delay_circle_orbit_a0}
\end{figure}

To illustrate how a deformation of the spacetime can affect our measurement, we simulate a simple case in which probe~A is on a circular orbit with radius $r_{\rm A}=50~M$ and probe~B is at a circular orbit with radius $r_{\rm B}=20~M$. First, we assume that the spacetime is described by the Schwarzschild solution. At coordinate time $t=0$, the $\phi$ coordinates of probe~A and probe~B are equal: $\phi_{\rm A}(t=0) = \phi_{\rm B}(t=0)$. Probe~B continuously emits an electromagnetic signal, which is later received by probe~A. We can solve photon trajectories with $n = -2$, $-1$, 0, 1, and 2 using Eq.~\eqref{eqn:emission_receiving_condition}. We define the proper time interval from emission at probe~B to reception at probe~A as $\Delta\tau_n$, where $n$ denotes the index in Eq.~\eqref{eqn:emission_receiving_condition}. Directly measuring $\Delta\tau_n$ requires knowledge of both probe~A and probe~B, which can be difficult and can introduce significant uncertainty. Instead, a neat way to extract $\Delta\tau_n$ is to measure the time difference between the arrival of photons with adjacent orders, $\Delta\tau_{n_1n_2}=\Delta\tau_{n_1}-\Delta\tau_{n_2}$, for $(n_1, n_2) = (-2,-1)$, $(-1,0)$, $(0,1)$, and $(1,2)$. This is feasible if signatures (such as the spectrum of the signal) can be encoded to help probe~A identify that these photons were emitted at the same instant. If we can measure this $\Delta\tau_{n_1n_2}$ with sufficient precision, the deformation of spacetime can be constrained. In Fig.~\ref{fig:time_delay_circle_orbit_a0}, we plot these time differences $\Delta\tau_{n_1n_2}$ as a function of the photon emission time $\tau_{\rm A}$, using the proper time of probe~A for both the emission time (x-axis) and the time difference (y-axis). First, these four time differences exhibit the same temporal evolution and differ only by a time translation. Second, every curve asymptotes to the time $\pm T_\gamma (1 - 3M/r_{\rm A})^{-1/2}$ (depicted by blue dashed lines in Fig.~\ref{fig:time_delay_circle_orbit_a0}), where $T_\gamma$ is the orbital period at the photon orbit measured in coordinate time $t$, and $(1 - 3M/r_{\rm A})^{-1/2}$ is the conversion factor from coordinate time $t$ to probe~A's proper time $\tau_{\rm A}$.

\begin{figure}[t]
\centering
\includegraphics[width=0.48\textwidth]{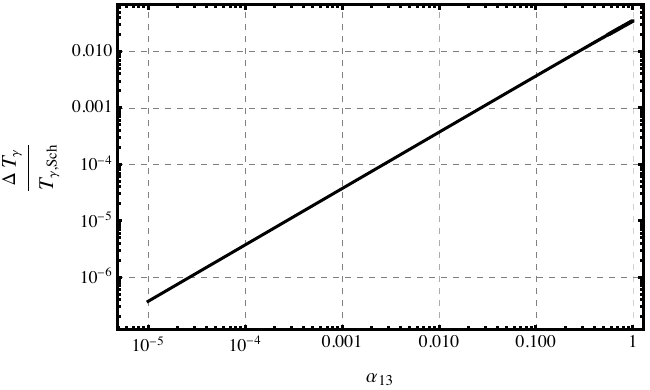}
\caption{Relative difference between the orbital period at the photon orbit in the Schwarzschild spacetime and that in the Johannsen spacetime as a function of the deformation parameter $\alpha_{13}$.}
\label{fig:dT_p_plot}
\end{figure}

In the Schwarzschild spacetime, $T_\gamma = 6\sqrt{3}\pi M$. In the case of a non-vanishing $\alpha_{13}$, the orbital period at the photon orbit is
\begin{equation}
T_\gamma = 2\pi \left( 1+\frac{\alpha_{13}M^3}{r_\gamma^3} \right) 
\left(1 - \frac{2M}{r_\gamma}\right)^{-1/2} r_\gamma \, ,
\end{equation}
which reduces to $T_\gamma = 6\sqrt{3}\pi M$ for $\alpha_{13} = 0$ and $r_\gamma = 3M$. Fig.~\ref{fig:dT_p_plot} shows the relative difference between the orbital period at the photon orbit in the Schwarzschild spacetime and that in the Johannsen spacetime. This quantity approximately follows a linear relationship versus the deformation parameter $\alpha_{13}$.

\begin{figure}[t]
\centering
\includegraphics[width=0.48\textwidth]{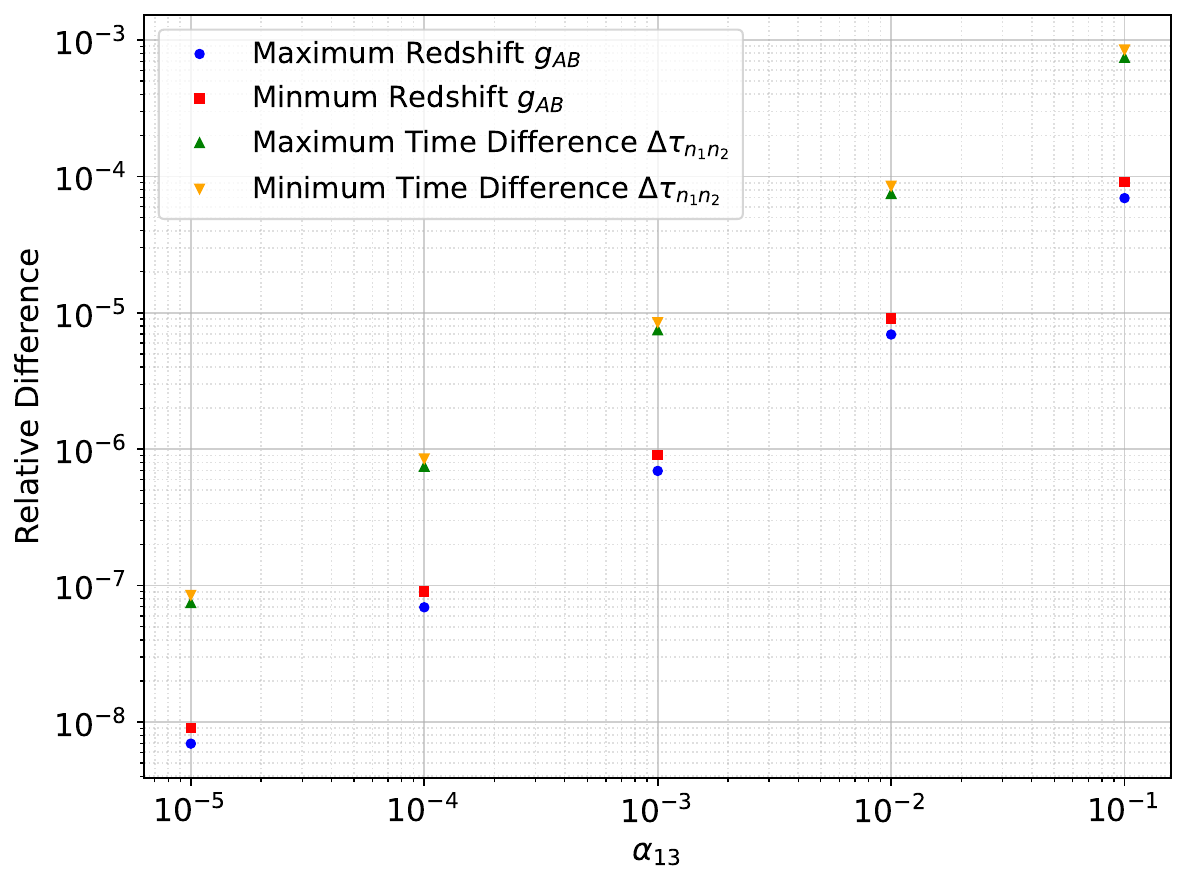}
\vspace{-0.6cm}
\caption{Relative difference of maximum/minimum redshift $g_{\rm AB}$ and the maximum/minimum time difference $\Delta\tau_{n_1n_2}$ for adjacent orders of $n$ between the Schwarzschild and Johannsen spacetimes.}
\label{fig:relative_difference_deformation_a13}
\end{figure}

We repeat our simulations for a non-vanishing deformation $\alpha_{13}$, and the results are summarized in Fig.~\ref{fig:relative_difference_deformation_a13}, which shows the relative difference of maximum/minimum redshift $g_{\rm AB}$ and the maximum/minimum time difference $\Delta\tau_{n_1n_2}$ for adjacent orders of $n$ between the Schwarzschild and Johannsen spacetimes. Here we define the relativistic redshift factor $g_{\rm AB}$ of the signal between the emitting probe B and receiving probe A as:
\begin{equation}\label{eqn:redshift_definition}
    g_{\rm AB} \equiv \frac{\nu_{\rm A}}{\nu_{\rm B}} = \frac{-u^\mu_{\rm A} k_\mu\big\vert_{\rm A}}{-u^\mu_{\rm B} k_\mu\big\vert_{\rm B}} = \frac{d\tau_{\rm B}}{d\tau_{\rm A}},
\end{equation}
where $u^\mu_{\rm B}$ and $u^\mu_{\rm A}$ denote the 4-velocities of the emitting and receiving probes (probe B and probe A, respectively), $k_\mu$ is the covariant 4-momentum of the photon evaluated at the emission ($\vert_{\rm B}$) and reception ($\vert_{\rm A}$) points, and $d\tau_{\rm B}$ and $d\tau_{\rm A}$ represent the corresponding proper time intervals. Fig.~\ref{fig:relative_difference_deformation_a13} provides a bottom-line estimate of the measurement accuracy required to constrain the deformation parameter $\alpha_{13}$ through measurements of the electromagnetic signal communication between the probes.

%%%%%%%%%%%%%%%%%%%%%%%%%%%%%%%%%%%%%%%%%%%%%%%%%%

\section{Testing the event horizon}\label{s-horizon}

Now we consider the case of a probe plunging into the black hole. When the probe is plunging, it continuously sends electromagnetic signals to the mothership in an outer orbit. In this configuration, we have the potential to test the strong field near the event horizon more accurately.

\begin{figure}[t]
\centering
\includegraphics[width=0.45\textwidth]{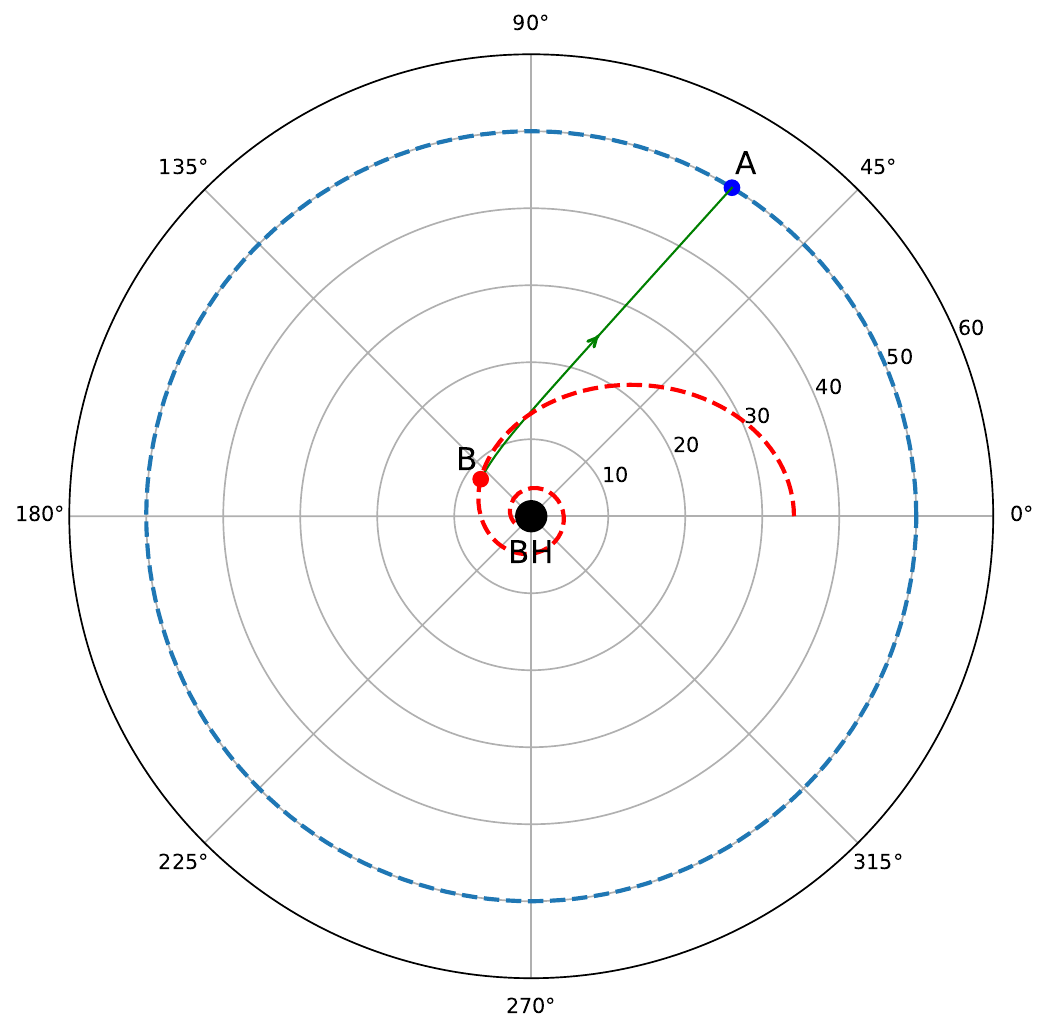}
\caption{Probe~B plunges onto a 10~$M_\odot$ Schwarzschild black hole and continuously sends electromagnetic signals to probe~A, which is in a circular orbit at $r_{\rm A} = 50~M$. All the data points are strictly obtained through numerical simulations. }
\label{fig:plunging_probe_config}
\end{figure}

We consider the system shown in Fig.~\ref{fig:plunging_probe_config}. The compact object is a 10~$M_\odot$ Schwarzschild black hole. Probe~A is in a circular at $r_{\rm A} = 50~M$. Probe~B is plunging onto the black hole (its trajectory is indicated by the red-dashed curve in Fig.~\ref{fig:plunging_probe_config}) and regularly emits a highly collimated signal (green curve with arrow), which is received by probe A.

\begin{figure}[t]
\centering
\includegraphics[width=0.45\textwidth]{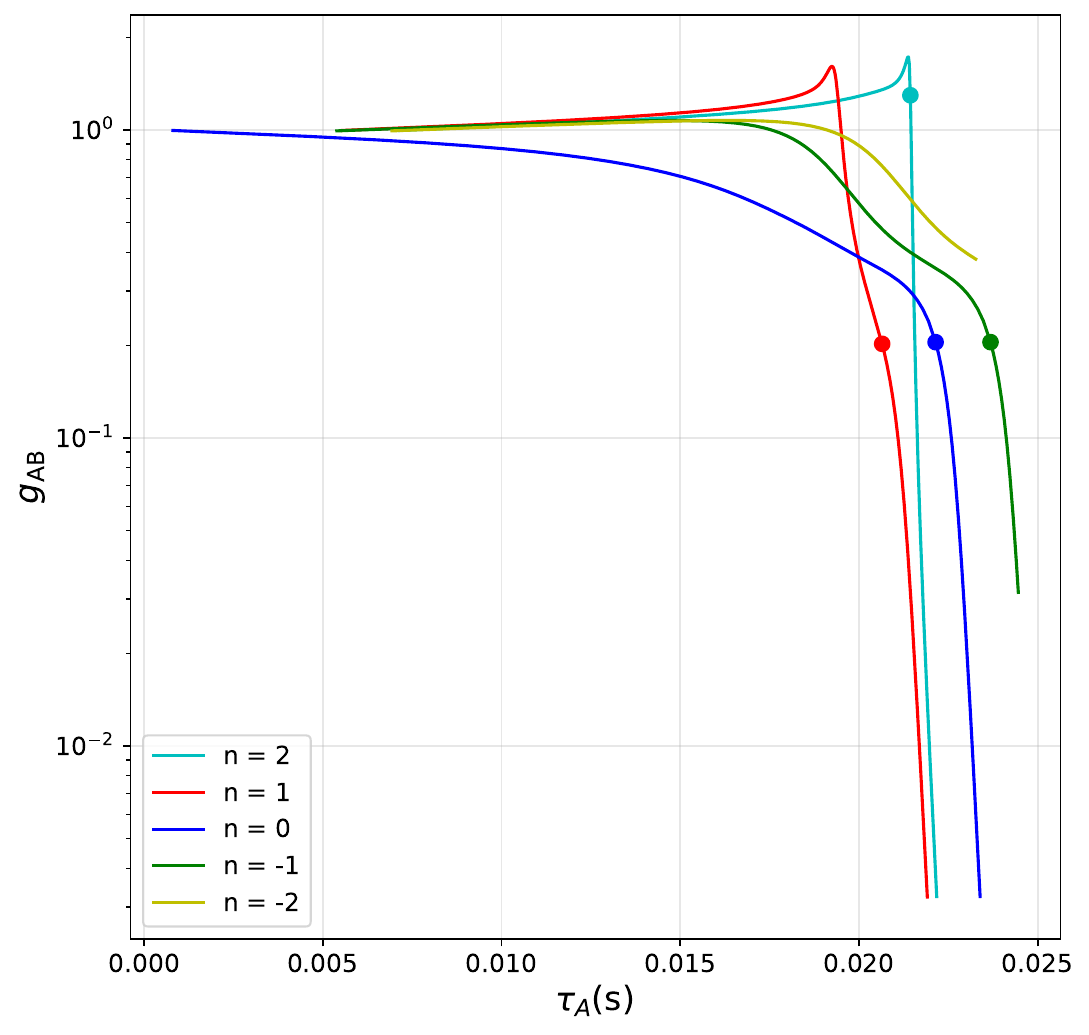}
\caption{Redshift $g_{\rm AB}$ of the signal emitted by probe~B and received by probe~A in Fig.~\ref{fig:plunging_probe_config} as a function of the proper time $\tau_{\rm A}$ of probe~A. The dots on the curves indicate the photons emitted when probe~B crosses the photon sphere.}
\label{fig:plunging_probe_redshift}
\end{figure}

In Fig.~\ref{fig:plunging_probe_redshift}, the redshift $g_{\rm AB}$ of the received signal is plotted versus the proper time of the receiving probe A, as defined by Eq.~\eqref{eqn:redshift_definition}. Here we computed the solutions to Eq.~\eqref{eqn:emission_receiving_condition} for $n=-2, -1, 0, 1, 2$ during the plunging of probe B, then calculated the redshift of the received photon at probe A. In Fig.~\ref{fig:plunging_probe_redshift}, the dots on the curves indicate the time when probe~A receives the photons emitted when probe~B crosses the photon sphere. In the simulation, the last signal is emitted by the probe B located at $r\approx2.012M$ close to the horizon. However, the local emission angle satisfying Eq.~\eqref{eqn:emission_receiving_condition} for $n=-1,-2$ is too stringent to find as $r\rightarrow2M$.  
Probe~B moves counterclockwise but the photon $n<0$ moves clockwise. When probe~B is close to the horizon, the emitted photon for $n=-1$ and $-2$ orbits more than two circles around the photon sphere. In such case, the outgoing direction of the photon ray is extremely sensitive to the initial emission angle due to the unstable nature of the photon sphere.

In general, the ability of testing the physics near the event horizon with a plunging probe depends highly on the ability of the highly collimated signal emitted by probe~B to reach probe~A and the frequency band of the detector on probe~A.

%%%%%%%%%%%%%%%%%%%%%%%%%%%%%%%%%%%%%%%%%%%%%%%%%%

\section{Concluding remarks}\label{s-conclusions}

Within 50~light-years of the Solar System, there may be a few stellar-mass black holes, even though no such candidate is known today~\cite{Nosirov:2026fjo}. The idea of sending spacecraft to a nearby black hole to test General Relativity in the strong-field regime is a fascinating possibility, despite its significant technological challenges~\cite{Bambi:2025kcr}.

In this manuscript, we have presented a preliminary study of how an interstellar mission with nanocrafts could test the spacetime geometry around a black hole, under the assumption that these nanocrafts have the capability to decelerate and enter certain orbits. In Paper~II~\cite{Fan:2026ule}, we will explore the case in which these probes cannot decelerate, and all scientific experiments must be performed during a flyby. Our study focuses solely on the measurements that could be performed to achieve our goals, assuming that the required instrumentation can be developed. Studying how to develop such instrumentation is also very important and can determine which of the measurements discussed in our work are feasible or not, but this is beyond the scope of the present study.

The takeaway message of our work is that the ability to decelerate the probes is not sufficient. To perform very stringent tests of the spacetime geometry around a black hole, we need at least one probe to orbit very close to the innermost stable circular orbit. In this regard, Tab.~\ref{tab:comparison_precession} succinctly summarizes our findings. If a probe orbits near the innermost stable circular orbit for a few hours, we can hope to measure a non-vanishing deformation parameter $\alpha_{13}$ as small as $10^{-5}$. If this is not the case (see the last two rows with $r_p = 1000~M \sim 15000$~km), the probe would need to orbit the compact object for a long time even to obtain relatively modest constraints on the spacetime geometry.

Finally, we remind the reader that the present work is a preliminary study with a number of simplifications. Throughout the manuscript, we have assumed that the black hole is non-rotating and in vacuum. A more accurate assessment of how to test the spacetime geometry around a black hole will require relaxing these assumptions.

%%%%%%%%%%%%%%%%%%%%%%%%%%%%%%%%%%%%%%%%%%%%%%%%%%

\section*{Acknowledgements}
%\section*{Funding}

This work was supported by the National Natural Science Foundation of China (NSFC), Grant No.~W2531002.

%\section*{Data Availability Statement}

%\section*{Code Availability Statement}

%%%%%%%%%%%%%%%%%%%%%%%%%%%%%%%%%%%%%%%%%%%%%%%%%%

\appendix 
\section{Locally Minkowskian reference frame of an orbiting probe}\label{app:local_frame}

To generate the initial conditions of the photon emitted from the probe, it is convenient to construct a local Minkowskian reference frame. If $u^\mu$ is the 4-velocity of the probe, $u^\theta=0$ since the motion of the probe is restricted to the equatorial plane. Because the orbit is not necessarily circular, the components $u^t$, $u^r$, and $u^\phi$ can all be non-zero. The 4-velocity $u^\mu$ is obtained directly from the high-precision simulation of the probe's trajectory.

\begin{figure}[t]
\centering
\includegraphics[width=0.45\textwidth,trim=0cm 7cm 0cm 7cm,clip]{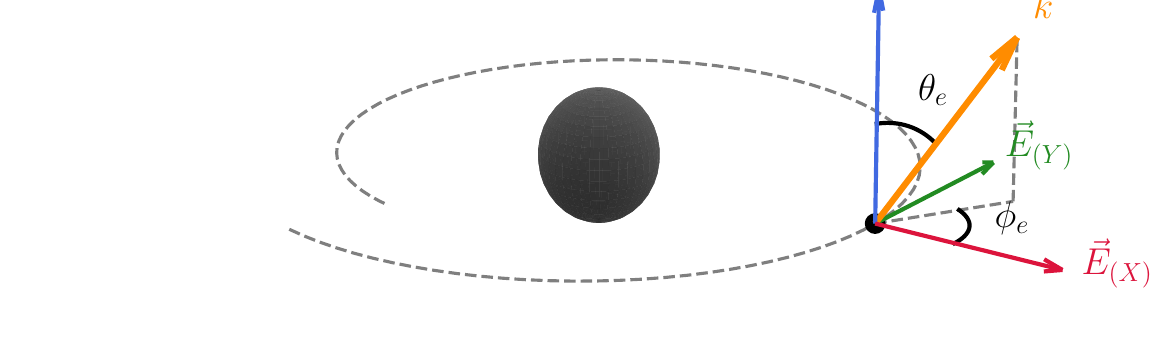}
\caption{Visualization of the local orthonormal spatial tetrad $(\vec{E}_{(X)}, \vec{E}_{(Y)}, \vec{E}_{(Z)})$ constructed along the probe's equatorial trajectory. The basis vectors are used to define the initial 4-momentum $\vec{k}$ of an emitted photon, parameterized by the local emission angles $\theta_{\rm e}$ and $\phi_{\rm e}$.}
\label{fig:local_tetrad_spherical}
\end{figure}

The time-like tetrad basis vector $E^\mu_{(T)}$ is defined as the 4-velocity $u^\mu$. We orient the space-like tetrad basis vectors as shown in Fig.~\ref{fig:local_tetrad_spherical}. For equatorial orbits, it is convenient to define $E^\mu_{(Z)}$ as
\begin{equation}
    E^\mu_{(Z)} = \frac{1}{\sqrt{g_{\theta\theta}}}\begin{pmatrix}
0 \\
0 \\
-1 \\
0
\end{pmatrix},
\end{equation}
where the negative sign ensures the spatial triad remains right-handed. By definition, $E^\mu_{(Z)}$ is strictly orthogonal to the equatorial motion. The remaining spatial basis vectors, $E^\mu_{(X)}$ and $E^\mu_{(Y)}$, must satisfy the standard orthonormality conditions: they must be unit spacelike vectors ($g_{\mu\nu}E^\mu_{(i)}E^\nu_{(i)}=1$) and mutually orthogonal to all other basis vectors ($g_{\mu\nu}E^\mu_{(i)}E^\nu_{(j)}=0$ for $i \neq j$).

To determine $E^\mu_{(X)}$ and $E^\mu_{(Y)}$, we employ the Gram-Schmidt orthogonalization process. We begin with a unit trial vector in the radial direction, $\tilde{E}^\mu_{(X)}=(0,1,0,0)^T$, and project out the time-like component:
\be
    E^\mu_{(X)} &=& \tilde{E}^\mu_{(X)}-\frac{(g_{\xi\nu}\tilde{E}^\xi_{(X)}E^\nu_{(T)})E^\mu_{(T)}}{g_{\xi\nu}E^\xi_{(T)}E^\nu_{(T)}} \nonumber\\
    &=& \tilde{E}^\mu_{(X)}+(g_{\xi\nu}\tilde{E}^\xi_{(X)}E^\nu_{(T)})E^\mu_{(T)}.
\ee
We then normalize $E^\mu_{(X)}$ to satisfy $g_{\mu\nu}E^\mu_{(X)}E^\nu_{(X)}=1$.

Next, using a unit trial vector in the azimuthal direction, $\tilde{E}^\mu_{(Y)}=(0,0,0,1)^T$, we obtain $E^\mu_{(Y)}$ by projecting out both the $E^\mu_{(T)}$ and $E^\mu_{(X)}$ components:
\be
    E^\mu_{(Y)} = \tilde{E}^\mu_{(Y)}+(g_{\xi\nu}\tilde{E}^\xi_{(Y)}E^\nu_{(T)})E^\mu_{(T)}-(g_{\xi\nu}\tilde{E}^\xi_{(Y)}E^\nu_{(X)})E^\mu_{(X)}. \nonumber\\
\ee
Again, we normalize $E^\mu_{(Y)}$ to satisfy $g_{\mu\nu}E^\mu_{(Y)}E^\nu_{(Y)}=1$.

In this local Minkowskian reference frame, the probe emits a photon with an initial 4-momentum $k^{(\alpha)}_0$ defined as
\begin{equation}
k^{(\alpha)}_0 =\begin{pmatrix}
k^{(T)}_0\\
k^{(X)}_0\\
k^{(Y)}_0\\
k^{(Z)}_0
\end{pmatrix}=\begin{pmatrix}
E\\
E\sin\theta_{\rm e}\cos\phi_{\rm e}\\
E\sin\theta_{\rm e}\sin\phi_{\rm e}\\
E\cos\theta_{\rm e}
\end{pmatrix},
\end{equation}
where $\theta_{\rm e} \in [0, \pi]$ and $\phi_{\rm e} \in [0, 2\pi)$ are the polar and azimuthal angles in the rest-frame of the probe, as illustrated in Fig.~\ref{fig:local_tetrad_spherical}. Finally, the photon 4-momentum in the global coordinate system $(t,r,\theta,\phi)$ is obtained via the coordinate transformation
\begin{equation}
k^\mu_0=E^\mu_{(\alpha)}k^{(\alpha)}_0.
\end{equation}

\bibliography{references}

\end{document}